\documentclass[conference]{IEEEtran}
\usepackage{cite}
\ifCLASSINFOpdf
  \usepackage[pdftex]{graphicx}
\else
  \usepackage[dvips]{graphicx}
\fi
\usepackage[hyphens]{url}
\usepackage{listings}
\usepackage{paralist} 
\usepackage{color}
\usepackage{booktabs}

\usepackage{xargs}
\newcommandx{\unsure}[2][1=]{}
\newcommandx{\change}[2][1=]{}
\newcommandx{\info}[2][1=]{}
\newcommandx{\improvement}[2][1=]{}
\newcommandx{\thiswillnotshow}[2][1=]{}

\begin{document}
\title{SONATA: Service Programming and Orchestration for Virtualized Software Networks}

%\author{\IEEEauthorblockN{Sevil Mehraghdam}
%\IEEEauthorblockA{University of Paderborn\\
%Paderborn, Germany\\
%sevil.mehraghdam@upb.de}
%\and
%\IEEEauthorblockN{Manuel Peuster}
%\IEEEauthorblockA{University of Paderborn\\
%Paderborn, Germany\\
%manuel.peuster@upb.de}
%\and
%\IEEEauthorblockN{Holger Karl}
%\IEEEauthorblockA{University of Paderborn\\
%Paderborn, Germany\\
%holger.karl@upb.de}
%}


\author{\IEEEauthorblockN{Sevil Dr\"axler\IEEEauthorrefmark{1},
Manuel Peuster\IEEEauthorrefmark{1},
Holger Karl\IEEEauthorrefmark{1}, 
Michael Bredel\IEEEauthorrefmark{2},
Johannes Lessmann\IEEEauthorrefmark{2},
Thomas Soenen\IEEEauthorrefmark{3},\\
Wouter Tavernier\IEEEauthorrefmark{3},
Sharon Mendel-Brin\IEEEauthorrefmark{4} and
George Xilouris\IEEEauthorrefmark{5}}

\IEEEauthorblockA{\IEEEauthorrefmark{1}Paderborn University: \{sevil.draexler, manuel.peuster, holger.karl\}@uni-paderborn.de}
\IEEEauthorblockA{\IEEEauthorrefmark{2}NEC: \{michael.bredel, johannes.lessmann\}@neclab.eu}
\IEEEauthorblockA{\IEEEauthorrefmark{3}UGent -- iMinds: \{thomas.soenen, wouter.tavernier\}@intec.ugent.be}
\IEEEauthorblockA{\IEEEauthorrefmark{4} NOKIA: \{sharon.mendel\}@nokia.com}
\IEEEauthorblockA{\IEEEauthorrefmark{5} NCSRD: \{xilouris\}@iit.demokritos.gr}
}


\maketitle

\begin{abstract}

In conventional large-scale networks, creation and management of network services
are costly and complex tasks that often consume a lot of resources, including time and manpower.
% Service creation and management are crucial processes in the competitive environment of network services. Due to the size and scale of the networks, however, these are non-trivial tasks that often consume a lot of resources, say man power, but still offer a limited quality of experience. Service creation, for instance, can often take hours or even days. Data centers, on the other hand, can set up compute services within minutes, if not seconds. 
Network softwarization and network function virtualization have been introduced to tackle these problems. They replace the hardware-based network service components and network control mechanisms with software components running on general-purpose hardware, aiming at decreasing costs and complexity of implementing new services, maintaining the implemented services, and managing available resources in service provisioning platforms and underlying infrastructures.
To experience the full potential of these approaches, innovative development support tools and service provisioning environments are needed. To answer these needs, we introduce the SONATA architecture, a service programming, orchestration, and management framework. We present a development toolchain for virtualized network services, fully integrated with a service platform and orchestration system. We motivate the modular and flexible architecture of our system and discuss its main components and features, such as function- and service-specific managers that allow fine-grained service management, slicing support to facilitate multi-tenancy, recursiveness for improved scalability, and full-featured DevOps support.

%EuCNC 2016 Deadline: 05. Feb, page limit 5 (6 if you pay 80eu)
%\vspace{6cm}

\end{abstract}



% \fi
\IEEEpeerreviewmaketitle

% put each main section into in own file (avoid merge problems when working in parallel)
\section{Introduction}
\label{sec:introduction}

%1st paragraph taken from D2.2
Service creation and management are crucial processes in the competitive environment of network services.
In conventional networks, service instantiation can often take several hours or even days. 
Also, managing the lifecycle of network services, including development, testing, resource allocation, deployment, scaling, monitoring, and debugging consists of a lot of expensive, inflexible, manual steps in hardware-based implementations.     
% lowering the quality of experience of customers and so the revenue of service providers. 
Consequently, there is a growing interest in network softwarization and network function virtualization (NFV), which aim to execute network service components, such as load balancers, firewalls, and deep packet inspectors as virtualized network functions (VNFs) on top of network infrastructures.

Taking into account the wide variety of services and service platforms with different requirements, the real potential of programmable, softwarized networks can only be utilized if flexible programming models, development support tools, management support tools, and execution environments 
are available. Addressing these challenges individually, however, is insufficient. Instead we need an integrated, consistent solution for the complete lifecycle of virtualized network services. Such a solution is still missing today.
%Existing efforts focusing on individual problems in this area fail to provide a flexible and 
To overcome this shortcoming, we introduce the SONATA architecture, a service programming
and orchestration framework that provides a development toolchain for
virtualized services, fully integrated with a service platform and orchestration 
system, designed and developed in the European project SONATA\cite{sonata.wepage}.

\begin{figure}[!t]
    \centering
    \includegraphics[width=1.0\linewidth]{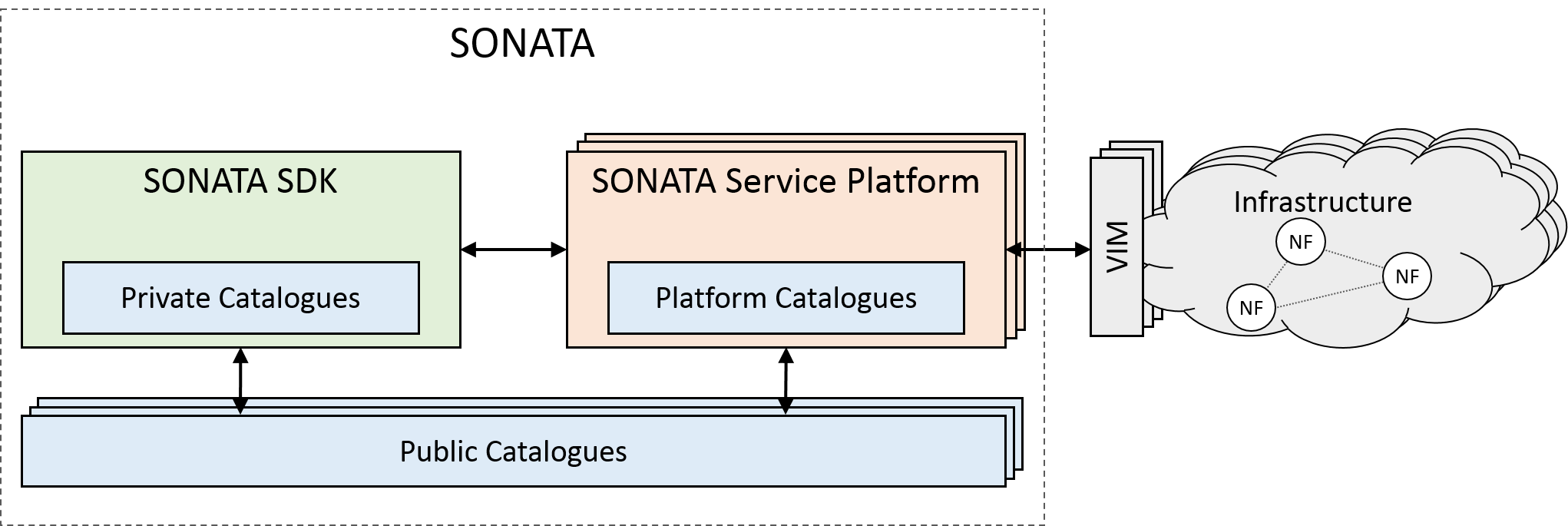}
    %\vspace{-0.5cm}
    \caption{Main architecture components of SONATA}
    \label{fig:arch-overview}
    % \vspace{-0.5cm} % trick to save some space. don't use if not needed.
\end{figure}

% first contribution
The main architectural components of SONATA are shown in Fig.\,\ref{fig:arch-overview}.
The first major component of SONATA is a Software Development Kit (SDK) that supports \emph{service developers} with both a programming model and a set of software tools. 
The SDK allows developers to define complex services consisting of multiple VNFs.
A \emph{service provider} (which might as well be the 
service developer) can then deploy and manage the services on one or more
SONATA service platforms.

% second contribution
SONATA's Service Platform (SP) is the second major component of the system,
offering a novel level of flexibility. Due to the fully customizable and
modular design of its management and orchestration framework, the SP offers customization opportunities on two levels.
First, the \emph{service platform operator} can modify the orchestration platform as such, e.g., to support a desired business model. Second, \emph{service developers} can influence the orchestration and management
functionalities of the platform pertaining to their own services, e.g., by including desired placement and scaling requirements in the service description, enabling the concept of \emph{Orchestration-as-a-Service} (OaaS). 
This empowers a new level of service control capabilities for service developers such as influencing placement decisions of services deployed across multiple points of presence (PoP). These PoPs can be full-fledged cloud data centres operated by \emph{infrastructure operators} but also smaller sites, like base stations that offer additional compute resources, e.g., in form of a couple of blade servers, operated by telco providers. 
Moreover, the service platform supports deployment and management of single services across different third-party infrastructures.

As shown in Fig.\,\ref{fig:arch-overview}, SONATA supports different catalogues storing artefacts like network functions and services, that can be produced, used, and managed by SONATA. Services developed and deployed by this system run on top of the underlying infrastructure. %accessible to the SONATA system via Virtual Infrastructure Managers (VIMs). 
The infrastructure needs to host and execute the actual network functions of a service, e.g., as a virtual machine. The service platform sends necessary information and instructions for execution and lifecycle management of services to the infrastructure.
The interaction between the service platform and the infrastructure is done through Virtual Infrastructure Managers (VIMs), e.g., OpenStack~\cite{OpenStack.wepage} or OpenVIM~\cite{OpenMANO.wepage}, which provide an abstract view on different infrastructure resources.

% third: DevOps through tight integration btw. SDK and SP
The tight integration between the SDK and the service platform bridges the gap between design and development of services on the one hand and deployment and lifecycle management of them on the other hand. Using the set of tools and techniques provided by the SONATA architecture, service developers and operators can collaborate to optimize the design and implementation of services, test and debug services using runtime information like monitoring or performance
data, and scale and adapt services to rapidly changing load and network resources.
This concept, known as a DevOps workflow in the software engineering community, supports agile service development and operation and takes 
the SONATA architecture beyond existing NFV platforms that focus on either the development or the operation side.

The rest of this paper is organized as follows. First, we present 
related work in Section~\ref{sec:related_work}. Section~\ref{sec:sdk} and Section~\ref{sec:service_platform} describe SONATA's SDK and service platform, respectively, and Section~\ref{sec:conclusion} concludes.

\section{Related Work}
\label{sec:related_work}

There are two categories of related work to be considered for SONATA's approach of applying agile software development methodologies to the field of NFV. The first category is about network programmability, programming models and tool support for the developer. The second category is about orchestration and management functionalities for service operation.

There is a variety of tools, like VeriFlow~\cite{khurshid2012veriflow}, and high-level languages, like Frenetic~\cite{foster2011frenetic}, that support the development process of SDN applications. Furthermore, the NetIDE~\cite{facca2013netide} project aims at providing a fully integrated development environment supporting the whole development lifecycle of SDN controller applications. However, all of them focus on the development of SDN control applications and do not consider generic network functions. The UNIFY project~\cite{csaszar2013unifying}, in contrast, gives first insights how to apply the DevOps model to NFV. They provide, for example, a multi-component debugging tool called Epoxide~\cite{levai2015epoxide}. Compared to these tools, SONATA's approach is a step forward and combines a powerful software development kit (SDK) with a flexible service platform to provide end-to-end support for service developers.

% In the area of orchestration and management tools much work already exists. The introduction of NFV paradigm into productions environment introduces a lot of challenges to be addressed (i.e., deployment and lifecycle management). 
In the category of Orchestration and management, solutions are mainly coming from the evolution of Cloud Orchestration platform like OpenStack~\cite{OpenStack.wepage} or OpenNebula~\cite{OpenNebula.wepage}, which provide the basic functionality to deploy and manage single predefined VMs but cannot handle composed services. Other solutions, like OpenStack Heat~\cite{OpenStackHeat.wepage}, Terraform~\cite{terraform.wepage}, and ADT~\cite{keller2013topology}, are able to deploy entire services composed of several VMs but do not focus on network function-specific needs, like flexible forwarding rules.  

%% This is maybe a too generic problem statement for this RW section. Removed for now, due to 5 page limit:
%
%As noted previously a key challenge is the scheduling of a service chain of VNFs across several Compute Nodes or OpenStack powered clouds. In order to address this challenge it is required to determine where the different network functions of a service chain are to be located, and decide an optimal method to scale those network functions. At this point  automated network optimization or resource management would require optimization of the entire network, including the legacy or non-virtualized component of the end-to-end service. Currently, the orchestrators are not always designed to handle this. The orchestrator, which is generic, needs to implement certain standardized interfaces towards VNF and OSS/BSS to provide automated solutions. 
The most notable approaches and projects directly focusing on NFV service orchestration can be divided into two categories. The first consists of research projects, like \mbox{T-NOVA}~\cite{xilouris2014t} and UNIFY~\cite{csaszar2013unifying,skoldstrom2014towards}, and the second category consists of open-source tools, like OpenMANO~\cite{OpenMANO.wepage}, OpenBaton~\cite{OpenBaton.webpage}, and OpenStack Tacker~\cite{Tacker.wepage}.
%, and the third category, which is out of the scope of this paper, includes commercial solutions provided by telco vendors, IT software providers and others.
There is also an upcoming first generation of NFVO commercial solutions, now appearing in the marketing of telecom vendors and solution providers. These solutions are often advertised as an extension to proprietary NFV management platforms and were not available for study beyond marketing material.

T-NOVA's and UNIFY's service platforms both provide basic orchestration functionalities for chained services described by a service graph. UNIFY's architecture aims at automated and dynamic service creation and recursive resource orchestration. Its orchestrator includes optimization algorithms for placement of service components; service-specific actions related to placement and scaling are deployed as a service component. T-NOVA via TeNOR~\cite{t-nova:tenor} is capable of orchestrating network services distributed across several data centers (NFVI-PoPs). T-NOVA supports chaining of multiple VNFs in each network service. Moreover, T-NOVA provides a VNF Marketplace for third-party VNFs. Currently the support for scaling is provided via the ingestion of rules and metrics declared in the VNF descriptors by the orchestrator.% for the definition NS scaling rules.
 
OpenMANO and OpenStack Tacker aim to be reference implementations of the MANO layer defined in the ETSI NFV ISG architecture~\cite{etsi-mano} but both are at the beginning of their development. Other academic orchestration solutions are Cloud4NFV~\cite{soares2014cloud4nfv} and vConductor~\cite{shen2015vconductor}. 

All presented orchestration tools, to a great extent, follow the same principle and try to build a single orchestration solution for different types of services. This creates multiple restrictions for service developers as they cannot influence the orchestration process as such. For example, some of the existing platforms allow expressing a limited and predefined set of preferences regarding monitoring, scaling, etc.\,within function descriptions. However, actively influencing service-specific decisions, e.g., placement and scaling of services and their components, is not supported by any of them. This is possible with SONATA's service platform using function- and service-specific manager programs defined by the service developer, as described in Section~\ref{subsubsec:FSM-SSM}.

\section{Software Development Kit}
\label{sec:sdk}
Support of DevOps, meaning the collaboration of software developers and other IT professionals to integrate and automate the software development and delivery and infrastructure changes, is vital to limit capital and operating expenses in today's communication networks. To this end, the SONATA project offers an SDK, which tightly integrates with the service platform and supports the development as well as the operation of network services. In this way, developers have access to monitoring data and performance measurements regarding services during the development phase, as well as the runtime. This information can be used for optimizing, modifying, and debugging the operation and functionality of services.

\subsection{Service Programming Model}
\label{subsection:programming_model}
To implement, describe, and deploy a network service that might comprise several network functions, SONATA uses various approaches. The overall service description is based on a domain-specific language, similar to TOSCA~\cite{tosca-nfv} and HOT~\cite{OpenStackHeat.wepage}, that defines the service components and their relationships. It allows users to describe deployments of complex network services in text files. These files are then parsed and executed by the SONATA service platform.
% We extend existing solutions by supporting declarative modeling of service-centric quality of service metrics, monitoring, and debugging.
%this was duplicated with the next paragraph! 
%We extend existing solutions by supporting declarative models for services that can describe quality of service metrics, monitoring, and debugging requirements. 
%We facilitate the use of service platform monitoring services, third-party monitoring services, and custom monitoring as part of the service description. 

The used modeling approach adds declarative descriptions for quality of service metrics, debugging requirements, and monitoring information supporting 
platform-specific, third-party, and custom monitoring services. By leveraging application-specific data acquisition mechanisms in addition to the underlying monitoring infrastructure, this enables service developers to specify monitoring metrics for collection and gain a comprehensive view of availability and performance. This information is exposed to other components of SONATA's service platform as well as the outside world, like external SDK modules and third-party applications, in a well-defined, authentication-protected way. The connected modules and applications can then process the information, analyze it, and react to it.

In addition, we support the description of function- and service-specific managers that allow service developers to customize the management of their services within the service platform.  With the service programming model, we can specify inputs and the expected behavior of these small management components. As a result, SONATA lets service developers choose fine-grained service control in a very flexible and agile way. Section~\ref{subsubsec:FSM-SSM} outlines this concept in more detail.

%\begin{itemize}
%	\item Explained the planned programming model.
%	\item Debugging, Monitoring, and Scaling
%	\item QoS requirements
%	\item SSM / FSM
%\end{itemize}

\subsection{Development Support Tools}
\label{subsection:development_tools}
SONATA's SDK design allows developers to define and test complex services consisting of multiple VNFs, with tools that facilitate custom implementations of individual VNFs. Fig.\,\ref{fig:sp} shows an overview of the SDK components. Developers create and maintain a SONATA workspace that contains all network service artifacts, such as description files, virtual machine images, and configuration files. Moreover, SONATA-specific editors support editing, verifying, and debugging service description files. The output from the editors is stored in the developer's workspace. These editors are also tightly coupled with the service platform and related service catalogs to store and retrieve artifacts and corresponding metadata. 

The package management tool uses the information stored in the workspace, like the service description, and creates service packages, similar to CSAR~\cite{tosca-csar} files. In SONATA we support slim packages that mainly contain references to artifacts as well as fat packages that can also contain large files like virtual machine images. 

The packages are stored in catalogs that can be within the SDK, within the service platform, or public. Given the right credentials, these catalogs can be accessed from the SDK and the data can be easily integrated into new network services. Moreover, packages can be deployed locally and executed on a container-based local service platform emulator.
%UPB: Don't refer to Dockernet for now. A paper for it is coming soon. :-)

SONATA's SDK also provides debugging and profiling tools that aim at shorter development cycles and facilitate the SONATA DevOps approach for network services. Using these tools, service developers can easily verify the performance and functionality of service components. The debugging tools help service developers to identify and eliminate errors within a given VNF or service. 
%They can trigger specific test patterns and automatically analyze the gathered results. 
Likewise, the profiling tools enable service developers to estimate resource and performance characteristics of a given VNF or service. They consist of several components, such as a traffic generator that generates a user-defined traffic workload, a monitoring agent that collects the different metrics as defined in the profiling description, and a profiling process that performs the actual test.

Finally, monitoring and data analysis tools, which interact closely with the service platform, expose the live service monitoring capability of SONATA to the service developer. These tools can operate on real-time data as well as historical data and help developers to understand and, if needed, debug their system.

%Say something about the SDK tools here. But let's keep it rather short, e.g., let's not explain the workspace structure or single son-* commands.
%
%\begin{itemize}
%	\item Packaging and verification
%	\item Catalogs
%	\item Monitoring
%\end{itemize}

\section{Service Platform}
\label{sec:service_platform}

\begin{figure*}[!t]
    \centering
    \includegraphics[width=1.0\textwidth]{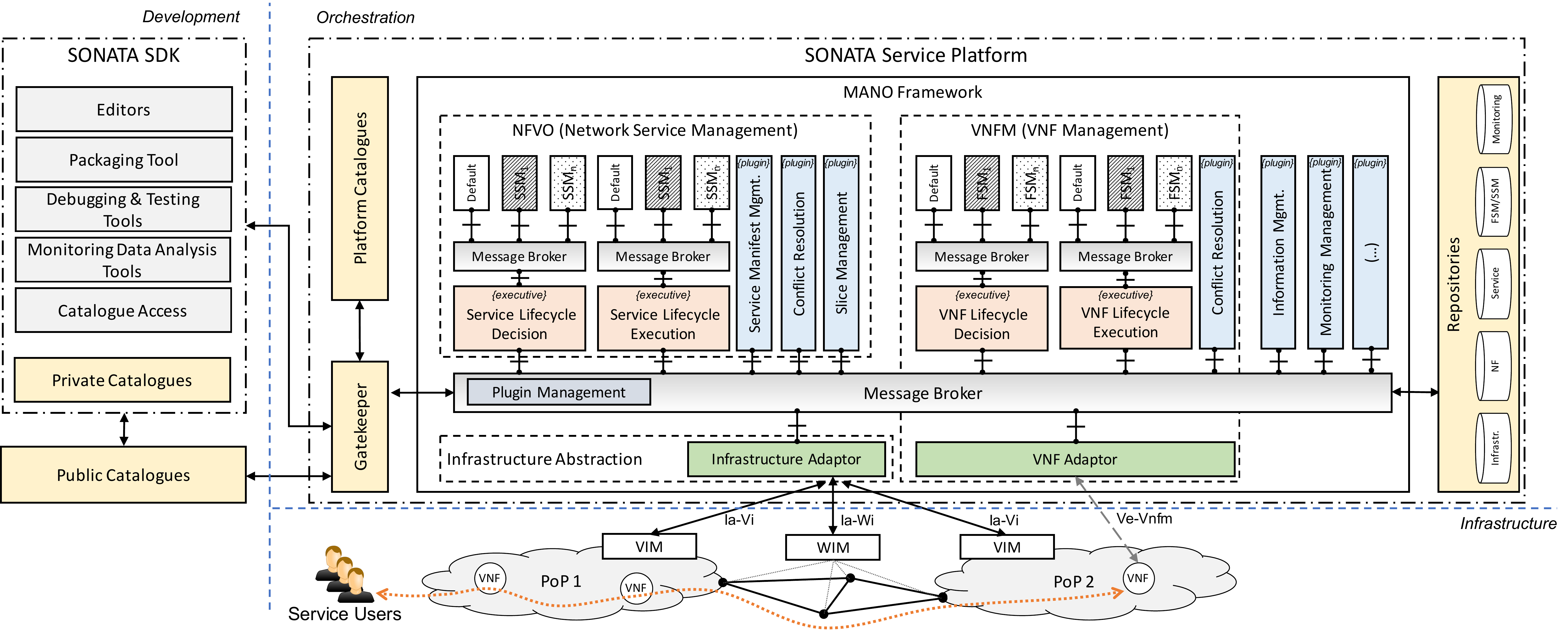}
    %\vspace{-0.5cm}
    \caption{SONATA's detailed architecture with SDK for service development, service platform for orchestration and management tasks and underlying infrastructure executing a network service.}
    \label{fig:sp}
    % \vspace{-0.5cm} % trick to save some space. don't use if not needed.
\end{figure*}

SONATA's service platform consists of four high-level components shown in Fig.~\ref{fig:sp}. The first component is the \emph{gatekeeper} module that is the main entry point of the service platform. It implements API endpoints for SDK tools, like the packaging tool, and it allows service developers to manually manage services deployed on the platform. It is also responsible to manage the access to service-specific monitoring information. The gatekeeper directly interfaces with the second platform component that is a \emph{platform-specific catalogue}, storing service artifacts uploaded to the platform. The third component contains \emph{repositories} for storing metadata of running services, e.g., monitoring data, placement results, and resource allocations. The last and main component is SONATA's extensible \emph{management and orchestration (MANO) framework} that implements the key functionalities of the platform and offers a novel level of management flexibility to both platform operators and service developers. Our platform utilizes existing cloud infrastructures and supports the management of infrastructure divided into multiple points of presence (PoP).

In the rest of this section, we describe the most important characteristics and components of 
SONATA's service platform.

\subsection{Customizable MANO Framework}

% genereal
The extensible MANO framework is the central entity of the platform and provides all functionalities to manage complex network services throughout their entire lifecycle. 
Almost every task that should be handled by the service platform, including the
service management and VNF management functionalities, is implemented using a set of loosely-coupled functional blocks, called \emph{MANO plugins} (Fig.\,\ref{fig:sp}). These plugins are connected via an asynchronous message broker that provides authenticated, reliable, and in-order publish/subscribe-based message delivery. Control communication among plugins takes place through this message broker. Other artifacts, like VM images, are maintained and shared with the help of common information bases, i.e., catalogues and repositories.

SONATA's service platform will include a set of default plugins that implement the basic management and orchestration behavior.
For example, we foresee default plugins for controlling the platform's monitoring capabilities or its information management, as well as adaptor plugins that connect to and abstract from a specific underlying infrastructure. By virtue of the asynchronous message broker, these plugins can be replaced easily (even at runtime) and new functionalities can be added into the platform by adding and integrating new plugins.
% plugins
% Using an asynchronous message broker makes the implementation and integration of new orchestration plugins much simpler than in architectures with classical plugin APIs. It also allows platform operators to replace plugins at runtime and easily integrate new functionalities into the platform by adding new plugins.
Owing to loose-coupling, we do not need to mandate a programming language for implementing these plugins. Similarly, we do not prescribe how the plugins are executed. For example, plugins may be executed as operating system-level processes, containers like Docker, or virtual machines within the platform operator's domain. It is even possible to execute MANO plugins remotely, which necessitates a secure channel between the remote site and the messaging system.
% , since all MANO plugins should be executed in an environment trusted by the platform operator.
%%% Removed the previous sentence as I coudn't relate to it as a ``reason'' for the first part of the sentence
The only requirement for a MANO plugin in this setup is the ability to communicate with the used messaging system. We give an overview of the communication patterns we foresee for the messaging system in Section~\ref{subsubsec:topic-based-comm}. 

% task of plugins /also plugins to outside (i.e. adators, monitoring)
% MANO plugins can implement almost every task that should be handled by the service platform. 
% For exmaple, we foresee  plugins for unpacking service packages uploaded to the platform, controlling the platform's monitoring capabilities, information management, as well as adaptor plugins that connect to and abstract from the underlying infrastructure. 
% SONATA's service platform will include a set of (replaceable) default plugins that implement the basic management and orchestration behavior. The set of plugins in the MANO framework can be extended as necessary.
% However, we do not fix this list, rather we predefine a list of default plugins to implement the basic management and orchestration behavior.

% SSM intro
In addition to the customizable plugin mechanism of the service platform,
% Another key feature of the MANO framework is the ability to customize 
the management and orchestration behaviour of the service platform with respect to individual network functions and services can also be customized. This is realized with \emph{function-specific managers (FSM)} and \emph{service-specific managers (SSM)} that are described in Section~\ref{subsubsec:FSM-SSM} in more detail.

\subsubsection{Topic-based Communication}
\label{subsubsec:topic-based-comm}

For the communication among MANO framework plugins, we outline a topic-based publish/subscribe
pattern that enables each component to talk to other components without the need to configure or announce API endpoints among them. This approach allows introducing additional components that are integrated into existing workflows without changing the component's implementation. This can either be done by reconfiguring existing components to emit messages to which the new component subscribes or by re-routing the messages on the message broker, e.g., by rewriting message topics. 

All components that want to connect to the system have to register themselves with a plugin manager that controls which messages are allowed to be sent and received by implementing a topic-based permission system. A hierarchical topic structure allows components to have fine-grained control over the information they want to receive. Topic subscriptions can be either specific, which means that a component subscribes to exactly one particular topic, or they can be generic by adding wildcard symbols to the subscription topic. The communication between MANO plugins are categorized into four top-level topics: \texttt{platform.*}, \texttt{infrastructure.*}, \texttt{service.*}, and \texttt{function.*}. Each of them is subdivided into further subtopics grouped by functionalities, like \texttt{*.management.*} or \texttt{*.monitoring.*}. Further extensions to this structure can easily be done by the platform operator by configuring the message broker accordingly and adding plugins that use the new topics.

\subsubsection{Function- and Service-Specific Managers}
\label{subsubsec:FSM-SSM}

FSMs and SSMs are small programs or workflow definitions implemented by a network function/service developer with the help of SONATA's SDK and shipped within the service package. Typical examples for such specific managers are custom service scaling and placement algorithms which place VNFs near to the users and automatically adapt the deployment to the current workload. Using these managers, the SONATA platform offers a novel level of flexibility to network function/service developers by adding programmability directly to the management and orchestration system. This goes beyond existing orchestration approaches where service management strategies are either limited to a predefined set of strategies or to simple, customizable rules, e.g., for autoscaling. FSMs/SSMs, in contrast, can be complete programs that can consume information like monitoring data, do complex computations to optimize their decisions, and instruct other components of the system to act accordingly.

Each MANO plugin in the system can allow such behaviour customizations by declaring itself as FSM-/SSM-capable. Such a customizable plugin is called an \emph{executive plugin} (Fig.~\ref{fig:sp}) and offers a northbound interface, based on a dedicated message broker, to interact with FSMs/SSMs that are integrated into the system on-the-fly when a new service package is uploaded. We call this procedure \emph{FSM/SSM on-boarding} and it includes validation procedures to check the messaging API of a FSM/SSM. In this design, the executive plugins are in charge of isolating FSMs/SSMs from the rest of the system and they can decide which information is accessible by each FSM/SSM. For example, an executive might modify substrate topologies used as inputs for placement SSMs to hide details of the network topology. Additionally, each executive plugin offers a number of FSMs/SSMs that implement the default behavior used in case a service package comes without its own FSMs/SSMs. Having a design with multiple, competing FSMs/SSMs will result in resource conflicts. 
SONATA will investigate conflict resolution solutions that regulate the decisions executive plugins take based on various FSMs and SSMs. Such a solution can be, e.g., based on game-theoretical auction mechanisms.
% We are aware of this issue and plan to provide conflict resolution mechanisms in our future work, which can, e.g., be based on game-theoretical auction mechanisms.

\subsection{Platform Recursiveness}

\begin{figure}[!ht]
	% \vspace{-0.2cm}
    \centering
    \includegraphics[width=1.0\columnwidth]{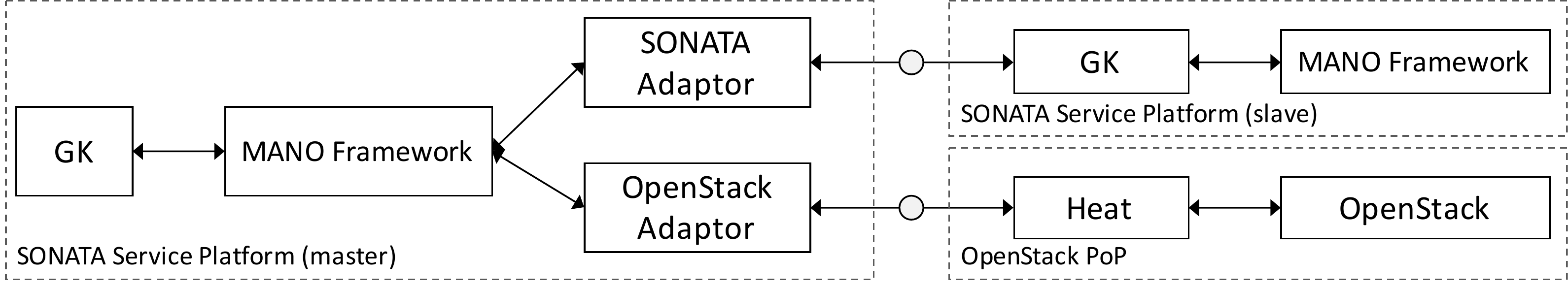}
    %\vspace{-0.5cm}
    \caption{Recursive deployment example: SONATA service platform managing a second service platform and an OpenStack instance}
    \label{fig:recursive}
    % \vspace{-0.5cm} % trick to save some space. don't use if not needed.
\end{figure}

SONATA's service platform implements a MANO layer running on top of an existing cloud infrastructure. A special case of this is a \emph{recursive platform deployment} in which a platform can delegate the management of services to underlying platforms~\cite{skoldstrom2014towards}. This improves scalability, as the same platform can be instantiated many times. The gatekeeper is the key component to enable such recursive deployments. It provides an interface to which other service platforms can connect and delegate deployment and management tasks. Fig.~\ref{fig:recursive} shows a recursive deployment in which the master platform manages a second service platform instance and an OpenStack-controlled PoP. One open challenge for these scenarios is automatically dividing complex service chains into smaller parts that can be deployed to and managed by different branches of such a recursive deployment.

\subsection{Slicing Support}
\label{sec:slicing}

The presented platform offers different options to support the concept of \emph{network slices} used to partition a physical network into a set of logically isolated networks. 
%Each of these slices has assigned a subset of available compute, storage, memory, and network resources enabling independent management and configuration of resources within a slice.
To implement this, our platform integrates an exchangeable slice manager plugin that can either use an external slice orchestration system or implement slice management functionalities by itself, e.g., by directly controlling SDN forwarding elements in the network. This approach allows platform operators to combine our system with almost every existing slicing solution.% offered by the underlying network infrastructure.

Based on the slicing concept, two service platform deployment models are possible. 
In the first model, a single SONATA service platform instance is responsible to do both the management of the network slices and the orchestration of services running inside these slices. We call this model \emph{flat slice management}. The second model, in contrast, utilizes the recursive deployment option of the platform. It uses one platform instance to manage the network slices and deploys an additional service platform instance within each of these slices. These additional platform instances are responsible for the orchestration of services within their particular slice. This concept should improve the overall system scalability and offers better isolation among services. We call it \emph{nested slice management}.

\subsection{Compliance with ETSI NFV Reference Architecture}
\label{sec:etsi}

\begin{figure}[!t]
    \centering
    \includegraphics[width=1.0\linewidth]{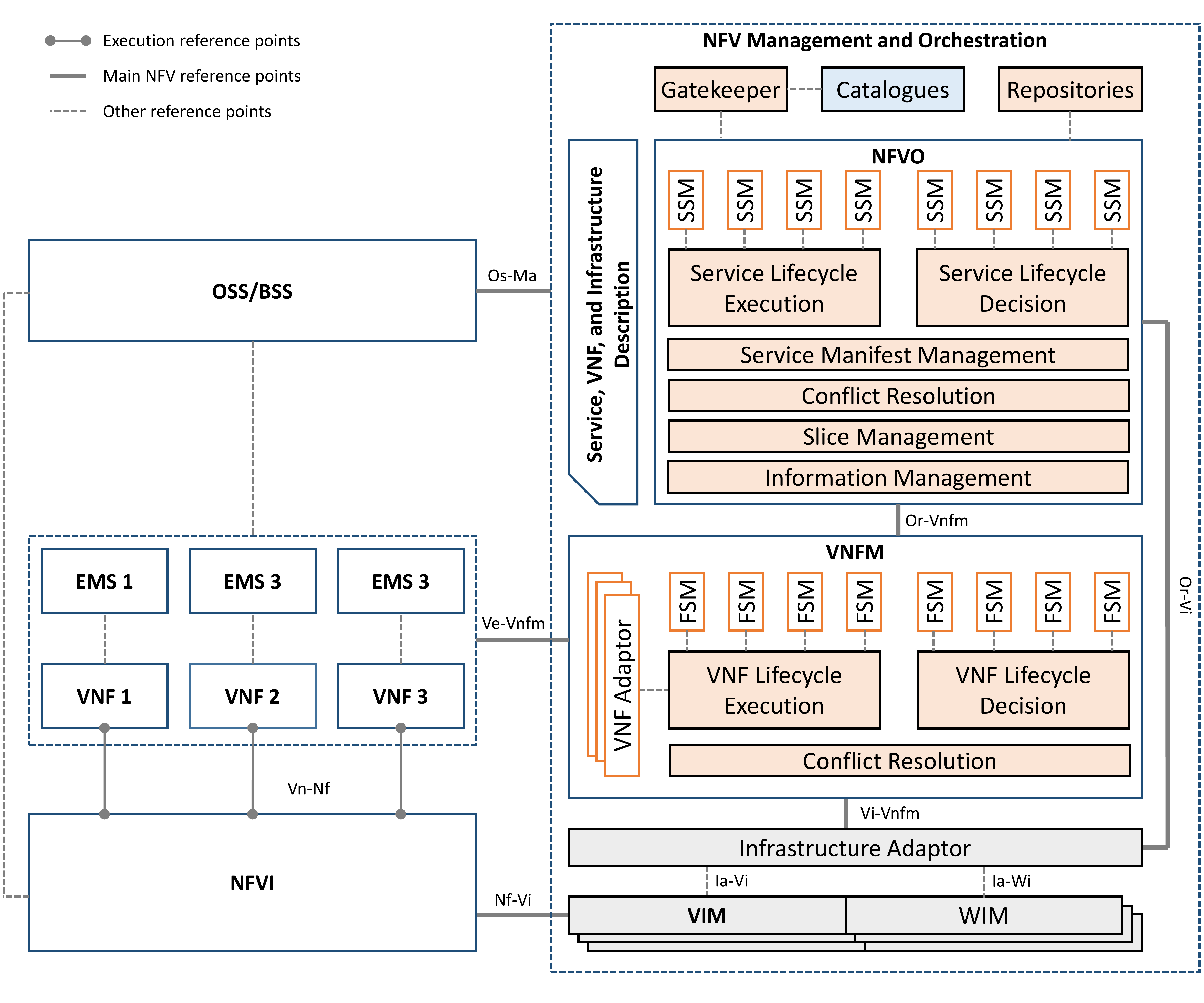}
    %\vspace{-0.5cm}
    \caption{Mapping functional architecture of SONATA to ETSI reference architecture.}
    \label{fig:etsi-mapping}
    % \vspace{-0.5cm} % trick to save some space. don't use if not needed.
\end{figure}

The functional architecture of the SONATA architecture complies with
and builds upon the ETSI reference architecture for NFV management and orchestration~\cite{etsi-mano}. 
As shown in Fig.~\ref{fig:etsi-mapping}, lifecycle management operations are divided into
service-level and function-level operations in SONATA. These operations include, for example, making decisions and executing decisions 
about placement, instantiation, scaling, and termination of network functions and network services. 
SONATA's service platform design defines the elements that build the NFV Orchestrator (NFVO) and VNF Manager (VNFM)
functionalities in ETSI's reference architecture. The key reference points of ETSI NFV are preserved (e.g., Or-Ma, Or-Vi, Ir-Vnfm, Vi-Vnfm interfaces) and complemented (e.g., Wi-Vnfm interface) in SONATA.

\section{Conclusion}
\label{sec:conclusion}

The high-level architecture design of the SONATA system
builds upon and extends the state of the art results and best practices proposed by industry, research
and development projects, as well as standardization bodies in the field of network function
virtualization and network softwarization. The proposed architecture provides a fully integrated development and deployment environment enabling the end-to-end support for network service development and operation. The described service platform provides a novel level of flexibility to both service platform operators as well as service developers and fully integrates into future 5G networks.
%%%% Removed this sentence as slicing is mentioned in the following sentence..
 % e.g, by supporting different slicing approaches. 

In particular, SONATA's architecture
provides a design for network slice management and multi-tenancy and supports recursive
installations of the service platform. SONATA offers full flexibility and convenience 
in adding and modifying management and orchestration functionalities on-the-fly, 
by virtue of a message broker system and loosely-coupled plugins, as well as 
function- and service-specific management programs. Tight integration between 
SONATA's software development toolkit and service platform and information exchange 
between these two main building blocks enables collaborative development and 
operation of complex network services.

Prototype implementations of the presented components and concepts validate the feasibility of  SONATA's architecture. A first release of the integrated SONATA SDK as well as SONATA's service platform will be open-sourced in the second half of 2016 by the SONATA consortium~\cite{sonata.wepage}.

% Using an agile development approach, the presented architecture design of the 
% SONATA system will be refined and amended as necessary based on detailed subsystem design and
% implementations.

\newpage

\section*{Acknowledgment}
This work has been performed in the framework of the SONATA project, funded by the European Commission under Grant number 671517 through the Horizon 2020 and 5G-PPP programs. The authors would like to acknowledge the contributions of their colleagues of the SONATA partner consortium (\url{www.sonata-nfv.eu}). This work was partially supported by the German Research Foundation (DFG) within the Collaborative Research Centre ``On-The-Fly Computing'' (SFB 901) and the International Graduate School ``Dynamic Intelligent Systems''.

\bibliographystyle{IEEEtran}
\bibliography{IEEEabrv,main}

\end{document}